# Survey on the State-of-the-Art in Device-to-Device Communication: A Resource Allocation Perspective

**Tariq Islam[1] and Cheolhyeon Kwon[1],**

[1]Mechanical Engineering Department, Ulsan National Institute of Science and Technology, Ulsan, Republic of Korea

Corresponding author: Cheolhyeon Kwon (e-mail: kwonc@ unist.ac.kr).

This work was supported by the National Research Foundation of Korea (NRF) Grants funded by the Korea Government (MSIT) under Grant 2020R1A5A8018822 and Grant 2020R1C1C1007323.

**ABSTRACT** Device to Device (D2D) communication takes advantage of the proximity between the communicating devices in order to achieve efficient resource utilization, improved throughput and energy efficiency, simultaneous serviceability and reduced latency. One of the main characteristics of D2D communication is reuse of the frequency resource in order to improve spectral efficiency of the system. Nevertheless, frequency reuse introduces significantly high interference levels thus necessitating efficient resource allocation algorithms that can enable simultaneous communication sessions through effective channel and/or power allocation. This survey paper presents a comprehensive investigation of the state-of-the-art resource allocation algorithms in D2D communication underlaying cellular networks. The surveyed algorithms are evaluated based on heterogeneous parameters which constitute the elementary features of a resource allocation algorithm in D2D paradigm. Additionally, in order to familiarize the readers with the basic design of the surveyed resource allocation algorithms, brief description of the mode of operation of each algorithm is presented. The surveyed algorithms are divided into four categories based on their technical doctrine i.e., conventional optimization based, Non-Orthogonal-Multiple-Access (NOMA) based, game theory based and machine learning based techniques. Towards the end, several open challenges are remarked as the future research directions in resource allocation for D2D communication.

## I. INTRODUCTION

The exponential growth in the number of wireless communication devices and related applications in the recent years has prompted dramatic rise in the wireless digital traffic. Thereof, new models have been studied to enhance the performance of the prevalent cellular framework in terms of connectivity and data rates [1]. Conventional solutions such as expanding the available spectrum, shrinking the cell coverage by employing femtocells or using multiple antennas are either not expedient in terms of cost due to the need for added infrastructure, or have already reached their saturation points [2]. On the other hand, Device-to-device (D2D) communication has been receiving ample attention lately due to its ability to dramatically improve simultaneous connectivity and data rates [3-6] which enables it to support a vast set of applications (See Fig.1). D2D communications facilitates direct communication between two or more devices with limited or no assistance from the core network [7]. The benefits of integrating D2D communication with the conventional cellular network include smaller communication delay due to direct communication, enhanced energy and spectral efficiency due to low power transmissions and spatial reuse of the spectrum respectively and traffic offloading [8]. In addition, D2D communication can be used to extend network coverage and boost the quality of experience of the edge users who generally suffer poor connectivity [9].

Device to Device communication has already been explored in 3$^{rd}$ Generation Partnership Project (3GPP) standardization as a proximity service (ProSe) that allows direct communication among devices situated in proximity. Specifically, the viability of D2D communication within the Long-Term Evolution (LTE) network and the different use cases are discussed in [10]. Besides, the architectural requirements for integrating D2D communication within the existing cellular network have been studied in [11]. In [12], the authors presented an overview of D2D fundamentals and standardization in 3GPP, including resource allocation.

Resource management is one of the crucial aspects of any D2D communication operation. In a broader context, resource management policies depend on the choice of spectrum for D2D communication i.e., unlicensed spectrum (outband D2D communication [13-17]) or licensed spectrum (inband D2D communication [7, 18-21]). Inband D2D communication can be further classified into underlay [22-25] and overlay [26-28] modes where the former refers to reusing the spectrum of Cellular Users (CUs) for D2D communication and the latter refers to dedicating a part of the licensed spectrum for D2D communication. There is a huge improvement in access rate, sum rate and spectrum efficiency in underlay mode as compared to overlay mode, but it is prone to induce high interference. Mitigating the effect of interference is a challenging issue which necessitates efficient power control and resource allocation strategies.

Recently, several literatures have been published to investigate and evaluate the resource management algorithms in D2D communication [29-36]. However, their scope is limited in certain areas. For instance, the surveys in [8, 31, 33, 34, 37] have limited coverage vis-à-vis the major underlying techniques for resource allocation i.e. Non-Orthogonal Multiple Access (NOMA) based, game theory based, machine learning based and conventional optimization-based resource allocation. Moreover, some of the surveys such as [30, 33, 34, 38] miss out on many of the recently proposed algorithms and/or the metric-based evaluation of the surveyed literature. In addition, some address only special cases such as Heterogeneous Network (HetNets) [31], millimeter-wave device to device networks [32] and socially aware D2D communication [39]. All in all, most of the existing survey articles are not far-reaching toward the latest breakthroughs and do not convey the algorithmic details, merits, and demerits of the individual algorithms in terms of quantifiable/qualifiable metrics.

In our view, tangible description of the different algorithms and their merits/demerits with respect to well-defined metrics is important to give the readers key insight on the current state-of-the-art. In this work, we extensively survey the recently proposed algorithms for resource and power allocation for D2D communication while systematically identifying their strengths and weaknesses. Specifically, our analysis takes into account a broad set of metrics which have direct impact on the performance of a resource and power allocation algorithm. In addition, brief explanation of the resource and power allocation strategy of each algorithm has been presented in order to acquaint the reader with the basic design of the algorithm. In our outlook the recent advances can be divided into four main categories based on the bottom-line techniques i.e., conventional optimization based [40-43], NOMA based [44, 45], game theory based [46-48] and machine learning based [49-52] resource allocation algorithms. (See Fig.2)

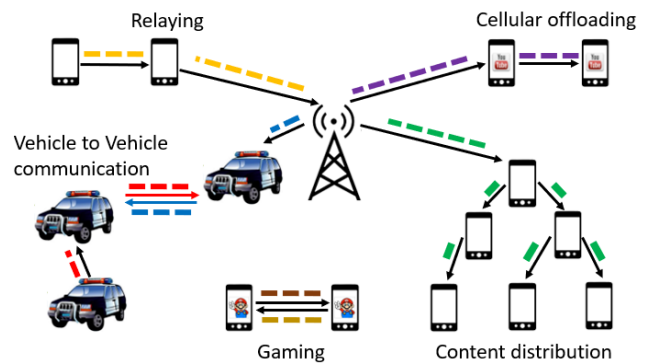

Fig. 1: Representative applications of D2D communication.

### A. PERFORMANCE OBJECTIVES

Following are some of the key performance objectives that are generally considered while designing a D2D resource allocation algorithm.

### 1) ACCESS RATE

Access rate refers to the number of D2D pairs that can be admitted to a given sub-channel/resource block. The access rate varies based on the algorithm design. For instance, algorithms such as [53-55] allow only one D2D pair per channel thus limiting the access rate to a single pair/group per channel. On the other hand, algorithms such as [56-58] allow multiple D2D groups per channel thus achieving significant improvement in access rate.

### 2) SUM RATE

Sum rate refers to the aggregate achievable data rate of all D2D and cellular users which depends on the achievable Signal to Interference and Noise Ratio (SINR) levels of the individual receivers. The SINR levels in turn depend on the overall interference levels. Efficient interference management and resource allocation lead to significant improvement in achievable sum rates.

### 3) FAIRNESS

Fairness, in the context of resource allocation, refers to the provision of equal resource allocation opportunity to all users irrespective of their position and required rates. Generally, access rate and sum rates can be improved by prioritizing D2D pairs with relatively low target SINRs. However, this results in an unfair system where D2D pairs with higher SINR requirements have lesser probability of being served.

### 4) RELIABILITY

Reliability refers to guaranteed and uninterrupted service to all the users. It is a crucial performance metric specially in applications such as Vehicle-to-Vehicle (V2V) communication, industrial automation etc. where unreliable connections may lead to highly fatal situations.

### B. EVALUATION METRICS

In this subsection, we briefly explain the evaluation metrics considered in this survey.

### 1) SPECTRUM SHARING MODE

Spectrum sharing mode can either be inband or outband where the former refers to sharing the licensed spectrum of cellular network for D2D communication and the latter uses unlicensed band. Inband sharing mode offers better spectrum utilization. However, Interference between cellular and D2D users necessitates careful resource allocation. On the other hand, interference between cellular and D2D users does not occur in case of outband communication but the interference from the devices operating in unlicensed band is generally uncontrollable.

### 2) UPLINK/DOWNLINK CHANNEL SHARING

Uplink channel sharing refers to sharing the uplink spectrum of cellular users for D2D communication. It simplifies interference management as interference occurs only at the Base Station (BS). On the other hand, in case of downlink resource sharing, interference occurs at multiple location thus making interference management and therefore resource allocation more complex.

### 3) NUMBER OF ALLOWERD D2D CONNECTIONS PER CHANNEL

The number of allowed D2D pairs per channel has a significant impact on access and sum rates. Allowing multiple D2D pairs per channel improves access rate which in turn improves sum rate as compared to only one D2D pair per channel.

### 4) CHANNEL ORTHOGONALITY

This metric determines whether the available channels are orthogonal or non-orthogonal. In case of orthogonal channels, the interference domain reduces to intra-channel interference due to the absence of inter-channel interference.

### 5) INTERCELLULAR INTERFERENCE

This metric determines whether the proposed algorithm considers inter cellular interference or not. Consideration of intercellular interference presents a more realistic scenario with additional challenges such as channel and power allocation to the edge users that are affected by intercellular interference. Nonetheless most of the existing literature considers only intracellular interference.

### 6) CHANNEL SHARING CONSIDERATION

The options that this metric can take are "one-to-one", "one-to-many", "many-to-many" and "many-to-one".

**One-to-one channel sharing:** One-to-one channel sharing refers to the scenarios where only one D2D pair is allowed per channel and no more than one channels can be allocated per user.

**One-to-many channel sharing:** One-to-many channel sharing refers to the scenarios where one D2D transmitter is allowed to transmit over multiple available channels. One-to-many channel sharing improves data rate by allowing transmitters to select multiple channels subject to the achievable overall data rate thus improving the probability of achieving higher throughput.

**Many-to-many channel sharing:** Many-to-many channel sharing refers to the situation when multiple D2D users are allowed per channel and multiple channels can be allocated to each user.

**Many-to-one channel sharing:** Many-to-one channel sharing refers to the situation when multiple D2D users are allowed per channel and each of the users can transmit only over one channel.

### 7) OPERATIONAL DESIGN

The operational design can be centralized, distributed or hybrid.

**Centralized design:** In case of centralized design resource allocation is administered by a centralized entity (such as base station) based on global Channel State Information (CSI). Centralized design offers better control and management at the cost of high communication overhead that is incurred to collect CSI.

**Distributed design:** In case of distributed design, resource allocation is administered locally based on the local CSI. The distributed design offers low communication overhead as only local CSI is required to make resource allocation decisions. However, it is relatively less efficient vis-à-vis resource allocation and interference management due to the absence of centralized control.

**Hybrid design:** Hybrid design combines both centralized and distributed approaches in order to achieve certain operational objectives.

### 8) TRANSMISSION MODE

A D2D group may consist of one transmitter and one receiver in which case the transmission mode is referred to as "unicast". Alternatively, a D2D group may consist of one transmitter and more than one receiver in which case the transmission mode is referred to as "multicast". Resource allocation in case of unicast is relatively simple as compared to multicast due to the existence of only one receiver per D2D pair.

### 9) QUALITY OF SERVICE (QoS) CONSIDERATION

This metric determines whether the QoS guarantees are provided for "only cellular users", "Only D2D users" or "Both cellular and D2D users".

In the metric evaluation tables henceforth, unless otherwise specified, the symbol "--" signifies that the corresponding metric can be either not determined from the available information or not applicable.

## II. D2D COMMUNICATION ALGORITHMS

In this section we list, categorize and analyze the D2D communication algorithms. The basic design, operations, merits and demerits of each algorithm are explained briefly. Moreover, each algorithm is evaluated based on a given set of metrics. Figure 2 shows the classification of the surveyed algorithms based on their bottom-line technique.

### A. CONVENTIONAL RESOURCE ALLOCATION ALGORITHMS

Typically, resource allocation problem in D2D underlaying cellular network scenarios is formulated as non-convex (e.g., continuous power allocation), combinatorial (e.g., user association and channel access), or Mixed-Integer Nonlinear Programming (MINP) (e.g., combination of continuous and discrete variables in optimization) problem. Numerous algorithms have been developed to solve such problems systematically and find either the global optimum solution or sub-optimal solution. These algorithms include, Weighted Minimum Mean Square Error (WMMSE), Fractional Programming (FP), genetic algorithms, among others. Such algorithms are highly computationally-demanding and generally implemented in a centralized manner. Moreover, they assume full and real time information about network statistics and CSI.

In this subsection, we survey some of the recent research on resource allocation in D2D communication using conventional methods. For better organization, this section is divided into two subsections based on the number of allowed D2D pairs/groups per channel. In the first subsection we review algorithms which allow single D2D pair/group per channel whereas the second subsection includes algorithms which allow multiple D2D pairs per channel. Table 1 and 2 present metric evaluation of the two categories respectively.

#### 1) SINGLE D2D PAIR/GROUP PER CHANNEL

Many of the conventional resource allocation algorithms allow channel sharing between a cellular user and only one D2D pair. This assumption simplifies interference management but results in lower access rates which in turn may result in lower sum rates. In the following text we have analyzed some of the recently proposed algorithms in this category.

In [59], the authors employ long duration time slots to enhance the system performance and fairness. The resource allocation problem is formulated considering proportional fairness among users where the sum of the proportional fairness functions of all the users is optimized. The optimization problem is resolved in two steps. The first step, i.e., system initialization phase, deals with first time slot and is divided into three sub stages; First, for all D2D and cellular users, the base station decides whether a given D2D pair can share spectrum with a given cellular user. Then joint power control is employed to allocate transmission powers optimally to cellular and D2D reuse partners subject to their SINR requirements and overall system throughput. Finally, aiming at optimizing the overall system throughput, Hungarian algorithm is employed for resource allocation. In step 2, which deals with time slot 2 and onwards, a joint transmission power control and proportional fairness scheduling method is proposed to maximize the sum of the proportional fairness functions of all the users. On the upside, the proposed algorithm achieves proportional fairness among users while maximizing system throughput. Moreover, the consideration of long time slot makes it a more practical solution for underlay D2D communication. However, despite considering vehicular communication, the channel model does not include parameters such as Doppler effect which is an important characterization factor for signal propagation between mobile objects

In [60] the authors address D2D based Vehicle-To-Everything (V2X) scenarios. Assuming imperfect CSI, the proposed algorithm aims at maximizing the sum ergodic capacity of the Vehicular User Equipment (VUEs) subject to the QoS requirements of the cellular and vehicular links. For power allocation, first the exact expression for ergodic capacity of single VUE is derived and then the simulated annealing (SA) algorithm [61] is used to solve the power control problem. Moreover, the computational complexity is reduced by using Jensen's inequality to simplify the exact expression of the ergodic capacity to an approximated closed form which is then solved by using the proposed power control method. Based on the results of the power control module, the original optimization problem is transformed to one to many resource allocation problem which is then solved using Gale Shapley algorithm [62]. The proposed algorithm is designed using imperfect CSI which is a realistic assumption ignored by many resource allocation algorithms. However, on the downside, the proposed algorithm assumes constant relative distance between the transmitter-receiver pair. Nonetheless, in practical scenarios the relative distance may change thus causing quick variations in the channel conditions between the transmitter and the receiver. This in turn may cause the allocated transmission power to be insufficient (e.g., in case of increasing distance between the communicating pair) and therefore may break the communication session thus compromising the performance of the proposed algorithm.

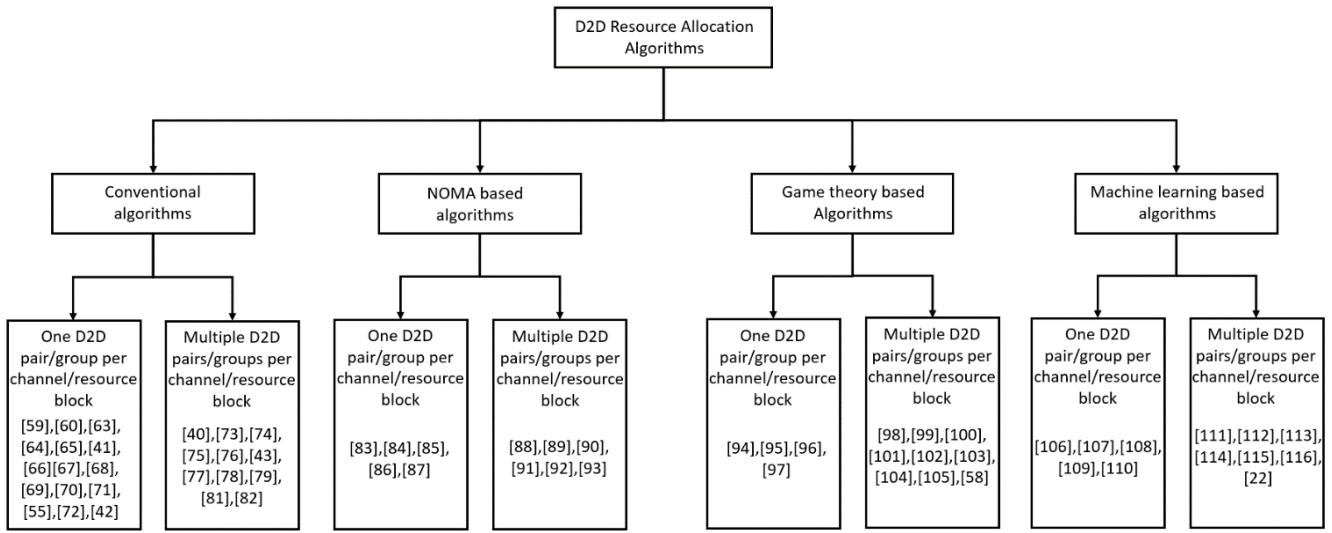

Fig 2. Classification of resource allocation algorithm for D2D communication

In [63] the objective function for power and resource allocation algorithm is formulated as aggregate power consumption minimization problem subject to the QoS requirements of the communication links in a one-to-many scenario where a D2D pair may reuse the assigned channels of multiple cellular users. Moreover, availability of partial CSI is assumed. Due to the NP-hard nature of the formulated problem, it is first converted into a tractable form through variable substitution and transformations. Then a two-layer greedy based outer approximation method is proposed to solve the transformed problem. In the proposed algorithm, the outer layer searches for near optimal number of activated reuse partners in each iteration and the inner layer estimates the objective power consumption. Moreover, a simplified version of the above mentioned two layers algorithm is proposed in order to deal with the scenario where uniform power is assigned to the D2D transmitter over each of the channels that it shares with its cellular reuse partners. The proposed two-layer method achieves near optimal resource allocation results. However, the assumption of D2D pair location only at the cell edges seems impractical and leaves room for evaluation of the proposed algorithm in more realistic scenarios where the D2D users may be located anywhere in the cell. Moreover, the stated assumption trivializes interference management (in single cell scenario as assumed in this algorithm) as users located on the cell edge generate minimal interference at the BS.

The algorithm proposed in [64] aims at maximizing the system sum rate while ensuring fair channel allocation to D2D users. The CSI is assumed to be imperfect and the channel uncertainties are tackled by probabilistic constraints that guarantee the desired outage probabilities. The resulting problem is formulated as a mixed integer non-convex optimization problem which is solved by decomposition into multiple power assignment subproblems and a resource allocation problem. Moreover, a closed form solution is obtained for power allocation subproblems in case of perfect CSI. However, in case of imperfect CSI, quadratic transformation and alternating optimization methods have been used. Both centralized and decentralized methods have been proposed to solve the formulated problem where in case of decentralized method, the computational load is distributed among the BS and the D2D pairs. The proposed algorithm achieves load balancing by distributing the computational complexity between BS and D2D users. Moreover, it reduces delays by allowing commencement of communication right after the first iteration. On the downside the proposed algorithm allows only one D2D pair per channel which results in underutilization of the available spectrum in terms of access rate and achievable data rates.

A two-step solution is proposed in [65]. In the first step, D2D and cellular links are assigned resource blocks based on channel gain information where the cellular and D2D users with least mutual interference channel gains are scheduled to reuse a given resource block. Subcarrier assignment is followed by power allocation in step 2 where cellular and D2D users are allocated transmission power subject to the power budget of the BS and harvested energy of D2D links. The merit of proposed algorithm lies in employing energy harvesting to boost up system performance. Its demerits include assuming D2D pair placement at the cell edges due to which the performance of the proposed algorithm in realistic scenarios where D2D pairs may be found anywhere in the cell has not been evaluated. Moreover, in a cellular environment, the edge users are prone to intercellular interference. This issue too has not been addressed as the authors assume single cell scenario.

In [41], aimed at improving network performance in terms of achieved sum rate, the authors propose a two-step algorithm in a multi cell environment where both intra and intercellular interference is taken into account. In the proposed algorithm, firstly admission control is performed which determines pairing between D2D and cellular users subject to their minimum QoS requirements. Then power control is applied to each D2D transmitter and its cellular reuse partner in order to maximize the network sum rate. The proposed method guarantees QoS for both cellular and D2D users in a realistic multicell scenario where it handles intra and intercellular interference effectively. On the downside, even though the proposed algorithm considers mobility in a multicellular scenario, it lacks appropriate channel model representation that includes mobility related aspects. Moreover, handover mechanism, which is an integral part of multicellular mobile communication, has not been addressed.

A two stage algorithm aimed at improving energy efficiency in one-to-one channel reuse scenarios is proposed in [66]. In the first stage, energy efficiency of D2D users is maximized subject to the QoS constraints of cellular and D2D users using Lamber W function which assigns optimal transmit power to a given D2D transmitter. Then, Gale-Shapley matching method is used in the 2nd stage to assign appropriate channels to D2D pairs. Overall, the proposed algorithm improves energy efficiency of the system while guaranteeing QoS for both cellular and D2D users. On the downside, the energy efficiency of the proposed algorithm is not scalable to the minimum SINR requirement of the cellular users and drops sharply with increase in the SINR threshold of the cellular users.

In [67] the authors observe that the parameters that effect the interference levels such as shadowing, topology and fading vary at different time scales. Based on this observation, the mode and channel allocation are handled centrally on a long-time scale while power allocation is carried out in a distributed manner where D2D terminals implement power control on a much smaller time scale using only local CSI and channel statistics. The long-term temporal design of mode and channel selection reduces signaling overhead by avoiding information collection across time. The signaling overhead is further reduced by making power control decisions based only on local CSI. The demerit of the proposed method lies in providing QoS guarantees only for cellular users.

An energy efficient resource, power and energy harvesting slot allocation problem is investigated for down link Energy Harvesting Heterogeneous D2D Networks (EH_DHNs) in [68]. The problem is formulated as an optimization problem aimed at maximizing energy efficiency. However, the formulated problem is a mixed integer nonlinear constraints optimization problem which is hard to solve. Therefore, it is transformed into a convex problem using variable relaxation method. Then in order to solve the simplified convex problem a joint energy harvesting time slot, resource block and transmission power allocation algorithm is proposed based on Lagrangian dual method and Dinkelbach method.

In [69], the proposed algorithm aims at guaranteeing reliability and latency requirements of V2V links. +The optimization problem is simplified by transforming the latency constraint into data rate constraint based on random network analysis using Poisson distribution to model packet arrival in the buffers of each V2V transmitter. Then a resource management algorithm is proposed. Keeping in view the computational intractability of the resource allocation problem, it is decomposed into independent sub problems using Lagrange dual decomposition and binary search methods. The complexity is further reduced through an iterative resource management algorithm where an optimal solution is derived for joint optimization of power and radio resource allocation with practical complexity.

Aiming at reducing latency, the authors in [70] propose a hybrid overlay resource allocation algorithm for V2V communication, where the V2V links are administered by the cellular base station. Based on the link establishment requests from Vehicles, the cellular BS chooses a receiver vehicle and allocates suitable channels subject to minimization of total weighted latency and SINR constraints. The formulated problem is comparable to Maximum Weighted Independent Set problem with Associated Weights (MWIS-AW). An analytical approach is adopted to compute the weights and the MWIS-AW problem is solved through a greedy cellular based V2V link selection algorithm.

The algorithm proposed in [71] allocates different channels to the potential receivers of a D2D group. Each receiver maintains a list of available channels and switches between the channels with a probability p which is determined by its dwelling time on a given channel (older nodes have higher probability of dwelling on a given channel and newer nodes have higher probability of switching). Thus, a cooperation among the nodes is formed and upon convergence, the nodes find their dwelling channels. Then a sender jump blind channel rendezvous algorithm is proposed which enables the senders to hop the available channels in order to rendezvous with their intended receivers. The proposed mechanism is scalable in the sense that it does not require intensive control information exchange between senders and receivers. Moreover, due to its distributed nature, it does not need to acquire global channel state information which is a non-trivial task due to the continuously changing channel conditions.

**Table 1: Metric evaluation (Conventional resource allocation algorithms for D2D communication: Single D2D pair per channel)**

| References | Inband/ Outband | Underlay/ overlay | Uplink/ Downlink | D2D pairs per channel | Inter channel interference | Inter cellular interference | Channel sharing mode | Operational Design | Unicast/ Multicast | QoS Guarantees |
|---|---|---|---|---|---|---|---|---|---|---|
| [59] | Inband | Underlay | Uplink | One | No | No | One to One | Centralized | Unicast | Both CU and D2D |
| [60] | Inband | Underlay | Uplink | One | No | No | One to many | Centralized | Multicast | CU (data rate) VUE (outage probability) |
| [63] | Inband | Underlay | Uplink | One | No | No | One to many | Centralized | Unicast | CU and D2D |
| [64] | Inband | Underlay | Uplink, downlink | One | No | Assumed to be constant and added with noise | One to many | Hybrid | Unicast | CU and D |
| [65] | Inband | Underlay | Downlink | One | No | No | One to many | Centralized | Unicast | Both CU and D2D |
| [41] | Inband | Underlay | Uplink | One | No | Yes | One to one | Centralized | Unicast | CU and D2D |
| [66] | Inband | Underlay | Uplink | One | No | No | One to One | Centralized | Unicast | CU and D2D |
| [67] | Inband | Underlay | Uplink | One | No | No | One to One | Hybrid | Unicast | CU |
| [68] | Inband | Underlay | Downlink | One | No | No | One to One | Centralized | Unicast | Only CU |
| [69] | Inband | Underlay | Uplink | One | No | No | One to many | Centralized | Broadcast | Only V2V |
| [70] | Inband | Overlay | -- | -- | No | -- | -- | Centralized | Can be unicast or multicast | V2V (CUs do not exist) |
| [71] | Can be Inband or Outband | Overlay in case of inband | Uplink | One | -- | No | One to One | Distributed | Unicast | None |
| [55] | Inband | Underlay | Uplink | One | No | No | One to One | Centralized | Unicast | CU and D2D |
| [72] | Inband | Underlay | Uplink | One | No | No | One to One | Centralized | Unicast | CU and D2D |
| [42] | Inband | Underlay | Uplink | -- | No | No | One to one | Centralized | Unicast | None |

A two-stage solution aiming at maximizing the overall throughput and minimizing power consumption and interference in multicellular environment is proposed in [55]. In the first stage resource allocation is performed such that the network throughput is maximized subject to the QoS requirements of D2D and cellular users while both inter and intra cellular interference are considered. Then in the 2nd stage dual Lagrangian approach is adopted for QoS aware optimal power allocation.

In [72], aiming at maximizing system fairness (vis-à-vis achievable data rates), first the proportional fairness scheduling algorithm is formulated. Then due to the NP-hard nature of the formulated problem, the original problem is decomposed into two subproblems i.e., optimal power allocation and channel assignment sub problems, which are solved sequentially. First the optimal power allocation sub problem is solved by transforming it into maximization of weighted sum of current data rates of all links. Then given the power allocation, the channel allocation subproblem becomes an Integer Linear Programming problem which is solved using linear programing.

In [42], the authors use fuzzy theory based algorithm for joint channel allocation and pairing. A set of D2D receivers is considered for reusing the cellular frequency resource. It is assumed that for each of the considered D2D receivers, there exists a set of D2D transmitter out of which the best D2D transmitter is selected using a stable fuzzy pairing criterion. The connection stability is ensured by assigning a fuzzy degree to each node where the fuzzy degree is a function of achievable data rates and battery levels of the corresponding nodes. The proposed method improves stability and fairness. However, QoS guarantees are not provided.

## 2) MULTIPLE D2D PAIRS/GROUPS PER CHANNEL

Some resource allocation algorithms allow channel sharing between a cellular user and multiple D2D pairs. This assumption complicates interference management. However, it produces higher access rates which in turn may result in higher sum rates. The following text investigates some of the recently proposed algorithms in this category.

In [40] a joint mode selection, power allocation and channel assignment algorithm is proposed, where mode selection refers to selection between licensed and unlicensed bands. The algorithm allocates continuous powers and discrete channels jointly to cellular and D2D users. In addition, multiple D2D pairs can be admitted to a given channel that is allocated to a cellular user. The D2D pairs that incur strong interference to the cellular users are identified using the Particle Swarm Optimization (PSO) algorithm. Such D2D pairs are pushed to the unlicensed band using duty cycle method in such a way that the minimum data rate requirement of the users in the unlicensed band i.e., WIFI users is not violated. Moreover, the rate constraints are guaranteed through suitable fitness values which prevent the algorithm from being trapped into infeasible solutions. The proposed algorithm provides QoS guarantees for D2D and cellular users over the licensed band. On the unlicensed band, it tries to maximizes data rates of D2D and cellular users while protecting the data rate requirements of the WIFI users. However, the ability to operate on the unlicensed band comes at the cost of additional hardware complexity in the form of network interfaces which can operate on the unlicensed band.

A distributed mode selection and power control algorithm is proposed in [73]. The proposed algorithm enables D2D nodes to acquire local information in real time and use it to make mode and power selection decisions subject to the communication quality of the D2D pair and the impact on the proximal ongoing uplink transmissions. Based on the local information, the D2D transmitter may adaptively choose between two modes i.e., direct D2D transmission when the interference is limited, or accessing the receiver through base station otherwise. Moreover, a relative capacity loss metric is proposed which measures the reduction in the capacity at the base station as a result of a D2D transmission. Based on the proposed metric, the base station may force a D2D transmitter to postpone its transmission. The proposed algorithm assumes the use of outband channel for control messages. However, the interference caused by other networks e.g., WIFI etc. in the outband control channel is not addressed.

In [74], aimed at maximizing system capacity by exploiting many-to-many channel allocation, the authors propose a two phase many to many channel allocation algorithm. In the first phase, each channel is assigned to exactly one D2D pair using Hungarian algorithm which maximizes the system capacity when each channel is occupied by exactly one user. Then in the second phase, more D2D pairs are admitted to each channel subject to their channel quality and the level of received interference from already admitted pairs. The proposed method attains near optimal performance despite its linear complexity. Moreover, it provides QoS guarantees for CU while QoS guarantees for D2D users are not provided.

In [75], the resource allocation problem is divided into channel assignment and power allocation subproblems. For channel assignment subproblem a heuristic algorithm is proposed which operates in two steps. Firstly, subject to the fulfilment of the minimum data rate of the resident cellular user, D2D pairs are assigned the channels that have high gains. Then to further improve the data rate, the D2D pairs are assigned more channels based on user selection criteria which dictates that the considered D2D pair should have acceptable channel quality on the corresponding channel and it should incur less interference to the users already assigned to that channel. After channel allocation, successive convex optimization is employed to transform the power allocation problem into a set of non-convex problems which are then solved iteratively to achieve efficient power allocation. The consideration of many-to-many channel sharing model in the proposed algorithm provides a number of possible resource sharing combinations for each D2D pair which improves the probability of D2D pair admission and results in achieving higher sum rates.

The authors in [76] propose a partially distributed factor graphs based joint pairing and power allocation algorithm where max-min message passing method is employed to maximize the least rate of all active nodes. In the proposed method, by limiting the role of the BS in pairing and power control, the NP-hard problem of D2D user pairing is reduced to a new distributed paradigm having manageable complexity. The proposed algorithm need only partial network information at the BS. Moreover, it is shown to outperform alternative methods in terms of complexity. Furthermore, consideration of only one channel leaves room for evaluation of the algorithm's performance in practical multi-channel scenarios. In such scenarios the proposed algorithm is expected to admit the most suitable pairs to the channels considered earlier thus leaving the less suitable pairs (e.g., the pairs located near the BS or the pairs with requirement of higher transmission power due to relatively bigger distance between the transmitter-receiver pair) for the channels considered later. This will result in performance degradation for the lately considered channels in terms of D2D admission rate and sum rate.

In [43], the authors adopt many-to-many resource sharing approach where they divide the formulated problem into two sub-problems for tractability; channel assignment and power allocation. In the channel assignment phase D2D groups are admitted to reuse certain channels subject to the interference power among D2D pairs sharing a given channel and aggregate gain in the system throughput. After channel assignment, Lagrangian dual optimization is employed to determine optimal power for each D2D transmitter subject to D2D sum rate maximization. Moreover, relationship between the transmission power of D2D transmitter and energy efficiency is studied via an energy efficiency maximization problem and a linear relationship is established between the two metrics. The proposed algorithm is expected to favor D2D pairs with relatively smaller distance between transmitter and receiver due to its policy of selecting those D2D pairs that cause lesser interference to the other receivers on the shared channel. Thus, the D2D pairs that require higher transmission power due to relatively bigger distances may be left out in scenarios where the number of D2D pair is much larger than the maximum capacity of available channels.

A semi distributed resource allocation algorithm is proposed in [77], where D2D users can take resource allocation decisions based on local channel information. In order to perform resource allocation, firstly devices populate their resource occupancy matrices through neighborhood information exchange. The occupancy matrix contains information about neighboring devices, available resource blocks and the channel quality index. Then, resource blocks are assigned based on a certain priority criterion. Due to its lesser dependability on the base station for resource allocation, the proposed method achieves quicker convergence and reduces the work load on the base station. However, on the downside, the proposed algorithm does not provide any quality-of-service guarantees.

Aimed at mitigating interference from D2D transmitter to the cellular user, the authors in [78] propose a mode selection and power control algorithm. Mode selection depends on a predefined channel gain threshold between D2D users where an aspirant D2D user operates in underlay or overlay mode if the gain is above the threshold. Otherwise, it operates in cellular mode. Moreover, the power control is dictated by the performance threshold of the cellular user i.e., the interference generated by the D2D transmitter should not violate the QoS requirements of the cellular user.

In [79], aimed at system capacity maximization, the authors propose algorithms for one-to-one, one-to-many and many-to-many channel sharing scenarios. In the one-to-one case, optimal bipartite matching is employed to remove the D2D pairs that may result in sum rate reduction if admitted. The proposed one-to-one algorithm is shown to achieve maximum sum rate for one-to-one scenario while guaranteeing the QoS requirements of both cellular and D2D users. Moreover, two variants of one-to-many case namely general and restricted sharing are proposed where the former achieves optimal sum rate with low admission rate. The latter approach builds on the general approach to improve the admission rate. Finally, inspired by [80], the many-to-many approach employs weighted graph coloring to improve both access rate and sum rate. The proposed algorithm improves access rate and sum rate by allowing admission of multiple D2D pairs per channel. Moreover, QoS guarantees are provided for both D2D and cellular users.

Aiming at minimize the energy cost, in [81] firstly a heuristic algorithm is proposed to allocate subcarrier to cellular and D2D users where D2D users that cause lesser interference and have good channel gains are admitted with higher priority. Then, utilizing the differences in the concave function structure of the problem constraints, the power allocation subproblem is transformed into a convex optimization problem which is then solved to achieve local optimal solution to power allocation.

Aiming at improving energy and spectral efficiency of underlay D2D multicast, the authors in [82] explore the tradeoff between the two metrics (i.e. energy and spectral efficiency) by formulating the energy efficiency maximization problem with constraints on spectral efficiency and maximum allowed transmission power. First a feasible power region is defined based on the outage thresholds of cellular and D2D transmissions. Then upper and lower bounds constraints on transmission power are derived as infimum and supremum vis a vis the feasible power region. Finally, the energy efficiency optimization problem with constrains on spectral efficiency and transmission power is formulated. However, the problem is non-convex and complex. Therefore, a suboptimal algorithm is defined where the maximum value of energy efficiency is defined as the sum of individual maximum energy efficiencies of K subchannels. Individual energy efficiencies are a function of the transmission power of the corresponding D2D transmitter over a given channel. Moreover, for the cases where the maximum allowed total transmission power constraint is violated, a systematic reduction in transmission powers of the transmitters is carried out such that the energy efficiency is maximized while conforming to the maximum total power constraint.

**Table 2: Metric evaluation (Conventional resource allocation algorithms for D2D communication: Multiple D2D pairs per channel)**

| References | Inband/ Outband | Underlay/ overlay | Uplink/ Downlink | D2D pairs per channel | Inter channel interference | Inter cellular interference | Channel sharing mode | Operational Design | Unicast/ Multicast | QoS Guarantees |
|---|---|---|---|---|---|---|---|---|---|---|
| [40] | Inband, outband | Underlay in case of inband | Uplink | Multiple | No | No | Many to One | Centralized | Unicast | CU and D2D |
| [73] | Inband | Underlay | Uplink | Multiple | No | Yes | Many to one | Distributed | Unicast | Only CU |
| [74] | Inband | Underlay | Uplink | Multiple | No | Yes | Many to Many | Centralized | Unicast | CU |
| [75] | Inband | Underlay | Uplink | Multiple | No | No | Many to Many | Centralized | Unicast | CU |
| [76] | Inband | Underlay | Uplink | Multiple | No | No | Many to one | Partially distributed | Unicast | CU and D2D |
| [43] | Inband | Underlay | downlink | Multiple | No | No | Many to Many | Centralized | Unicast | CU |
| [77] | Inband | Underlay | Uplink | Multiple | No | No | -- | Semi-Distributed | Unicast | None |
| [78] | Inband | Underlay, overlay | Uplink | Multiple | No | No | One to One | Centralized | Unicast | CU |
| [79] | Inband | Underlay | Downlink | Multiple | No | No | One to One, One to Many, Many to Many | Centralized | Unicast | CU and D2D |
| [81] | Inband | Underlay | Uplink | Multiple | No | No | Many to Many | Centralized | Unicast | Both CUs and D2D |
| [82] | Inband | Underlay | Uplink | Multiple | No | No | Many to many | Centralized | Multicast | Both CU and D2D |

### B. NON-ORTHOGONAL-MULTIPLE-ACCESS BASED ALGORITHMS

NOMA can improve spectral efficiency by enabling transmitters to transmit multiple signals over the same channel using power superposition. As shown in Fig. 3, combining NOMA with D2D communication enhances channel sharing among multiple users by employing Successive Interference Cancellation (SIC) which exploits the difference in the signal strengths of the signals received from different sources in order to decode the intended message. Attracted by the potential improvement in spectral efficiency and massive connectivity offered by the integration of D2D communication with NOMA technology, many recent researches have proposed integration of D2D communication with NOMA technology.

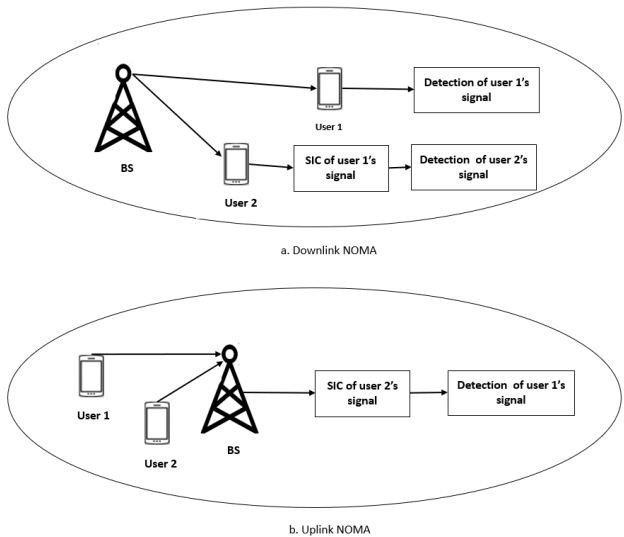

**Fig 3. General representation of NOMA technology**

#### 1) SINGLE D2D PAIR/GROUP PER CHANNEL

In [83], aiming at maximizing the energy efficiency of cellular users and NOMA based D2D users, the authors propose a distributed power and spectrum allocation algorithm for uplink NOMA based D2D communication in energy harvesting setup. A two-stage game is proposed to solve the energy efficiency maximization problem. In the first stage, using the proposed approximation algorithm the power allocation between potential cellular and D2D reuse partners is modeled as a non-cooperative game where cellular users and D2D groups act as players. Building on the equilibrium obtained in the first stage, a matching game is used in the second stage to get a global stable matching result. The use of non-cooperative game model in the first stage reduces the signaling overhead and computational complexity. Moreover, the computational complexity is further reduced through an energy aware screening method for the establishment of preference lists.

In [84], aiming at improving energy efficiency using energy harvesting, the authors investigate a joint time allocation and power control problem for D2D communication underlaying NOMA based cellular networks where D2D and cellular users harvest energy in the downlink from a hybrid access point for a certain time duration $\tau_e$ and then transmit data to their respective receivers in the uplink for time $\tau_i$. An iterative method is devised for maximizing the energy efficiency subject to the quality-of-service constraints of the cellular users. The proposed algorithm derives the global optimum solution in closed for by analyzing Karush–Kuhn–Tucker conditions in each iteration. However, on the downside, it considers a very simplistic scenario of only one aspirant D2D pair. Realistic scenarios with multiple aspirant D2D pairs present a variety of challenges in terms of interference management, and power and resource allocation which must be addressed.

In [85], the authors propose a transmission power optimization algorithm for NOMA-based D2D systems with imperfect successive interference cancellation decoding. Specifically, an optimization problem has been formulated to maximize the throughput of D2D users subject to the QoS requirement of the cellular user and the transmission powers of D2D transmitters. The formulated power allocation problem is non-convex in nature and therefore is first transformed into convex form. The problem is then solved using dual theory where the dual variables are updated using sub-gradient method. The proposed mechanism shows significant improvement in the data rates. Nevertheless, NOMA is employed only for D2D communication. The performance can be further improved by applying NOMA for the cellular pairs as well.

In [86], the authors investigate joint optimization of power control, channel allocation and mode selection (underlay mode or interlay mode). The interlay mode is proposed to improve successive interference management at the BS by multiplexing D2D and cellular users in the power domain. The optimization problem is formulated as a maximum weight clique problem in graph theory where the vertex weights are calculated using an iterative power allocation method based on sequential convex approximation. Moreover, Optimal solution for channel assignment and mode selection is found through a branch and bound based method which iteratively eliminates vertices corresponding to inefficient allocation. In the proposed algorithm, the power domain multiplexing of cellular and D2D pairs enabled by the proposed interlay mode effectively eradicates strong interference between the cellular and D2D users which in turn improves the sum rate.

Table 3: Metric evaluation (NOMA based resource allocation algorithms for D2D communication: One D2D pair per channel)

| References | Inband/ Outband | Underlay/ overlay | Uplink/ Downlink | D2D pairs per channel | Inter channel interference | Inter cellular interference | Channel sharing mode | Operational Design | Unicast/ Multicast | QoS Guarantees |
|---|---|---|---|---|---|---|---|---|---|---|
| [83] | Inband | Underlay | Uplink | One D2D group with multiple transmitters | No | No | One to one (One cluster of D2D transmitters over one channel) | Distributed | Multiple transmitters, one receiver | Both CU and D2D |
| [84] | Inband | Underlay | Uplink | One | Yes | No | One to one | Centralized | Unicast | Only CU |
| [85] | Inband | Underlay | Downlink | One | -- | No | One to One | Centralized | Multicast | CU |
| [86] | Inband | Underlay, interlay | Uplink | One in underlay mode | No | No | Many to One (cellular), One to one (D2D in underlay mode) Many to One (D2D in interlay mode) | Centralized | Unicast | CU and D2D |
| [87] | Inband | Underlay | Downlink | One | No | No | Many to One (in case of CU), One to One for D2D | Centralized | Unicast | CU and D2D |

Aimed at maximizing the sum mean opinion scores, the authors in [87] proposed a quality-of-experience based NOMA-enabled channel and power allocation algorithm. Optimal power allocation and channel assignment is determined through a two-step iterative algorithm that is based on constrained concave-convex technique and alternating optimization. The proposed method achieves significant improvement in QoE performance. However, the QoE is measured vis-à-vis only web browsing whereas more demanding application such as gaming, video sharing etc. which form a major demand area for D2D communication have not been considered.

### 2) MULTIPLE D2D PAIRS/GROUPS PER CHANNEL

In [88], firstly, NOMA-based cellular users and Non-NOMA D2D users are clustered to share a given channel where the clustering problem is formulated as a sequentially operated matching game with externalities. Then complementary geometric programming is employed to solve for power allocation. Finally, an iterative algorithm is proposed for joint optimization of the two objectives i.e., transmission power allocation and user clustering. The proposed algorithm has been shown to improve connectivity and provide interference protection to cellular users. Nonetheless no QoS guarantees have been provided in terms of data rate.

In [89], aiming at interference mitigation and throughput maximization, the proposed algorithm addresses channel allocation and group association through a 3-dimensional matching game between D2D transmitters, D2D mobile groups and subcarriers. Then the power of D2D mobile users is optimized over each subcarrier using brand and bound method. Moreover, successive convex approximation low complexity technique is employed to reduce the complexity of the proposed algorithms.

In [90] a two-step algorithm is proposed where, firstly, D2D users are grouped into clusters where NOMA is employed for many to one communication. Then, for easier tractability, the resource allocation problem is decomposed into subcarrier assignment and power allocation subproblems where the former subproblem is solved through many-to-many matching game and the latter subproblem is solved through a genetic algorithm-based optimization technique. A stable solution is obtained by iterative execution of the algorithms proposed for the two subproblems. In the proposed algorithm, consideration of social relationship among users enables realistic resource allocation and achieves higher sum rate while providing QoS guarantees to both cellular and D2D users.

In [91], firstly D2D mobile groups are formed to mitigate intra user interference through successive interference cancellation. Then resource allocation algorithm for cellular and D2D mobile users is designed aiming at mitigating cochannel and cross channel interference. Moreover, a channel reuse algorithm for channel sharing among D2D mobile groups is designed based on group rate selection criteria. Finally, power optimization is achieved through difference of two convex function approach. The proposed algorithm considers mobility. However, it does not take into account the variations in the channel state due to mobility which may have significant impact on the quality of the received signal which in turn may require increasing or decreasing the transmission power.

A NOMA based solution is proposed in [92]. The formulated problem is a non-convex problem and therefore it is simplified by splitting it into two subproblem; power allocation to receivers in D2D groups and subcarrier assignment. The sub-carrier assignment subproblem is addressed through many-to-one matching theory. Based on the results of subcarrier assignment, the solution to the non-convex power allocation subproblem is found using sequential convex programming that is shown to be convergent. The proposed algorithm is shown to approach close to the performance of exhaustive search algorithm with tolerable complexity. Moreover, NOMA based D2D communication improves sum rate and access rates as compared to its orthogonal counterpart.

Aiming at maximizing the sum rate of users in femtocell based network, the authors in [93] propose an algorithm for user pairing where NOMA interference is mitigated by pairing weak and strong users. Moreover, in order to maintain the quality of service for both macro and femto users a greedy based channel allocation algorithm and a dynamic power allocation method is proposed. The proposed algorithm achieves improvement in sum rate for femto users as compared to conventional cognitive radio orthogonal multiple access techniques. Moreover, it provides QoS guarantees for femto cell users.

### C. GAME THEORY BASED ALGORITHMS

Game theory is typically employed for distributed resource management in D2D communication scenarios where the players such as D2D users and base stations compete or cooperate to attain access to the radio resources. The resource management problem can be modeled as a non-cooperative or cooperative game between BSs and D2D users. In a cooperative game, players solve the primary resource allocation game collaboratively using optimization-based or heuristic methods to accomplish a certain goal such as spectral efficiency or sum rate maximization. In case of non-cooperative game on the other hand, players act in a noncollaborative and greedy manner in order to attain their own goal such as achieving their own QoS demands. The key objective of most of the game theory-based algorithms is to achieve the Nash Equilibrium for the underlying resource allocation problem.

### 1) Single D2D pair/group per channel

In [94], aimed at spectral efficiency improvement and overhead reduction, the authors propose a two-time scale resource allocation algorithm where D2D users assist in QoS improvement of the cellular users by acting as relays. In the proposed algorithm, the pairing between D2D and cellular users is determined at a long-time scale where the pairing problem is formulated as a distributed matching game-based approach and solved through a distributed algorithm which converges to ϵ-stable matching. At a short time-scale, an optimal cooperation policy allots the transmission time for cellular users and D2D pairs based on instantaneous channel state information. The proposed algorithm reduces signaling overhead significantly by using statistical CSI for pairing process whereas for deciding the transmission time it requires instantaneous CSI only between the matching pairs. However, on the downside, the QoS guarantees are provided only for cellular users.

**Table 4: Metric evaluation (NOMA based resource allocation algorithms for D2D communication: Multiple D2D pairs per channel)**

| References | Inband/ Outband | Underlay/ overlay | Uplink/ Downlink | D2D pairs per channel | Inter channel interference | Inter cellular interference | Channel sharing mode | Operational Design | Unicast/ Multicast | QoS Guarantees |
|---|---|---|---|---|---|---|---|---|---|---|
| [88] | Inband | Underlay | Downlink | Multiple | Yes | No | Many to One | Distributed clustering, Centralized power allocation | Unicast | Interference protection guarantees for CU |
| [89] | Inband | Underlay | Uplink | Multiple | No | No | Many to Many | Distributed | Unicast | CU and D2D |
| [90] | Inband | Underlay | uplink | Multiple | No | No | Many to one | Distributed | Unicast | CU and D2D |
| [91] | Inband | Underlay | Uplink | Multiple | No | No | Many to Many | Centralized | Multicast | CU and D2D |
| [92] | Inband | Underlay | Uplink | Multiple | No | No | Many to one | Centralized | Multicast | CU and D2D |
| [93] | Inband | Underlay | Downlink | One Cognitive weak femto user and one cognitive strong femto user | Yes | Yes (femto cells) | One to One | Centralized | Unicast | Both macro and femto users |

In [95], the resource allocation problem is modeled with the help of Generalized Nash Bargaining Solution (GNBS) framework where the cellular users decide about sharing resources with the D2D users. The resource sharing aims at maximizing a payoff function that captures the system sum rate by introducing a bargaining factor. The formulated problem is NP-hard and therefore divided into two subproblems i.e., Channel assignment and power allocation. Channel assignment is solved through the max-weighted max flow algorithm whereas Lagrangian multiplier method is employed to solve power control. The proposed algorithm achieves better data rates as compared to its baseline algorithms and provides QoS guarantees for both cellular and D2D users.

In [96], a two stage Stackelberg game is formulated where the Cellular user and the NOMA-enabled D2D users act as leader and followers respectively. In the first stage, Kuhn-Munkers algorithm is employed to pair up cellular and D2D users as resource reuse partners i.e., D2D users are allocated subcarriers of their partner cellular users. Then in the second stage particle swarm optimization is used for optimal power allocation to D2D users. The proposed method uses social relationship parameters to strengthen the cooperation between cellular and D2D reuse partners and achieve efficient resource allocation.

In [97] a Transferable Utility game play is proposed for ensuring fairness among users and providing QoS guarantees. The resource reuse problem is modeled as a Stackelberg game that determines optimal transmission power of the users and thereby enables resource reuse by controlling the mutual interference among the resource reuse partners. The proposed algorithm scales well due to its polynomial computational complexity, provides QoS guarantees and ensures fairness among user terminals.

### 2) MULTIPLE D2D PAIR/GROUP PER CHANNEL

A fully distributed algorithm for joint channel and power allocation to D2D pairs in time varying environment subject to the quality-of-service requirements of cellular users is proposed in [98]. The problem is modeled as a Stackelberg game with pricing. In each iteration of the game, the base station broadcasts the set of available resource blocks and their corresponding prices. This information is used by the D2D pairs to learn the transmit power levels and usable channel through a stochastic learning algorithm where each player (i.e., D2D pairs) derives its strategy by observing its own payoffs generated by its actions. Then, the price vector is updated and rebroadcasted by the base station for guaranteeing the quality-of-service requirements of the cellular users. Using the updated prices, the D2D pairs relearn the transmit power levels and usable resource blocks through the stochastic learning algorithm. The proposed framework is less computationally extensive, has nominal signaling overhead and does not require global channel state information. However, it does not provide any QoS guarantees on D2D communication.

Aiming at improving the Quality of Experience (QoE) of the users and provisioning resources based on application category, the resource allocation problem in [99] is modeled as a dynamic Stackelberg game that employs multiple utility functions for flexible and efficient application driven resource allocation. In order to enhance their QoE, D2D users are divided into three groups based on their application class. For each application class a different utility function is defined. Moreover, as the utility functions corresponding to different application classes are different, they can't be merged into a single utility function using nonlinear or linear scalarization. To solve this, a distributed algorithm is proposed based on Stackelberg game with non-scalarized multiple criteria optimization which is shown to converge to a unique sub-game perfect Stackelberg Equilibrium. In the proposed algorithm, the base station and D2D pairs play leader and follower game respectively where the BS manages the intra cellular interference by imposing a fee on D2D users for resource reuse. The D2D pairs respond by selecting optimal power levels and suitable resource blocks. The proposed algorithm mitigates interference through efficient transmit power reduction and thereby improves throughput, while guaranteeing quality of experience for all D2D pairs.

In [100], aiming at maximizing energy efficiency a two-stage algorithm is proposed. In the first stage, subject to the rate and power constraints, a coalitional game with a partition function based on the combined co-channel interference is employed to solve channel allocation subproblem. Here overlapping coalitions are permitted in order to expand the solution space. The proposed distributed method achieves stable but suboptimal resource allocation. For power control, fractional programing is employed in the 2nd stage where optimal transmission power is determined for each transmitter in a centralized manner. The two stages are executed in an alternate way in order to optimize the transmission powers for each coalition. The combined framework allows maximizing individual and system level energy efficiency.

In [101], the authors propose a game theory-based downlink resource block sharing algorithm for intercellular D2D communication. A two-player infinite horizon repeated game is designed where an inter or intra cellular D2D pair is considered as one player while its neighboring base stations jointly are considered as the other player. A D2D pair plays repeated game with the set of its neighboring base stations in a way that the players maximize their utility function by sharing a subset of their initially allocated set of resource blocks with the other player. In order to maximize the utility of all the players, cooperation between the players is enforced by penalizing the non-cooperating players. The punishment for the non-cooperative player is in the form of throughput reduction which is enforced by enabling the opponent player to reduce the number of its shared resource blocks with the non-cooperating player. However, in order to achieve quick convergence to the equilibrium the punishment period is reduced by lowering the value of the penalty factor. The proposed algorithm assumes a relatively more realistic scenario by considering a multi cellular environment and achieves significant improvement in average throughput.

Table 5: Metric evaluation (Game theory-based resource allocation algorithms for D2D communication: One D2D pair per channel)

| References | Inband/ Outband | Underlay/ overlay | Uplink/ Downlink | D2D pairs per channel | Inter channel interference | Inter cellular interference | Channel sharing mode | Operational Design | Unicast/ Multicast | QoS Guarantees |
|---|---|---|---|---|---|---|---|---|---|---|
| [94] | Inband | Underlay | Uplink | One | -- | No | One to one | Distributed, | Unicast | CU |
| [95] | Inband | Underlay | Downlink | One | No | No | One to One | Centralized | Unicast | CU and D2D |
| [96] | Inband | Underlay | Uplink | One | No | No | One to One | Distributed | Multicast | CU and D2D |
| [97] | Inband | Underlay | Uplink | One | No | No | One to One (one CU & one D2D user) Two to one (Two D2D users) | Distributed | Unicast | Both CU and D2D |

A D2D resource allocation algorithm for D2D communication underlaying heterogeneous cellular networks comprising millimeter wave and traditional cellular communication is proposed in [102]. The resource allocation problem is modeled as a coalition game with transferrable utilities where each player i.e., D2D pair is assigned to either millimeter wave band or a subchannel in the licensed cellular band subject to the corresponding improvement in the overall sum rate of the system. The proposed algorithm converges to a Nash equilibrium and reaches a near-optimal solution at a fast convergence rate. However, on the downside, the proposed algorithm considers interference among only its own D2D pairs on the unlicensed mm-wave band while ignoring interference from the other networks operating on the same band.

The authors in [103] propose a fully uncoupled learning algorithm where the D2D users learn pareto optimal actions vis-à-vis resource block and transmission power level selection. The QoS of the cellular users is ensured by modeling the resource allocation problem as Stackelberg game where the BS acts as leader which is responsible for price updates of the resource block and the D2D pairs act as followers who, using the pricing information, learn pareto optimal actions to maximize the D2D sum rate.

In [104], the authors propose a robust distributed power allocation and nonuniform price negotiation algorithm for V2V communication underlaying cellular networks where the communication channel is characterized by uncertain channel conditions and co-channel interference. Because of the channel uncertainty, the interference, delivery rate and delay related constraints are presented in probabilistic form which make the problem non-convex and intractable. Therefore, Bernstein approximation is employed to convert it into a solvable closed form. The resource and power allocation problem is modeled as a non-cooperative Stackelberg game with a single leader (i.e., the base station) and multiple-followers (i.e. the V2V pairs) where power control is achieved through a price-penalty mechanism. The proposed price-penalty mechanism discourages selfish actions and balances the benefits for all users. The authors, propose their solution considering realistic scenarios with pronounced effect of mobility on channel condition of the interference channel. However, for the signal channel between communicating pair, the channel gain is assumed constant which is unrealistic because of the multiple practical factors such as the relative speed of the communicating vehicles, variation in fading due to active highway environment etc.

In [105], the authors propose game theory-based frameworks for distributed power allocation for overlay and underlay D2D communication in cellular networks. In case of the overlay mode, the objective function of each user is divided into two terms i.e., a log term that caters for the throughput of the corresponding user and a linear term which is interpreted as the penalty for exploiting a particular resource. In case of underlay mode, an optimum solution is found for the power allocation problem which is formulated with an added constraint on the transmission power of the D2D users so that the regular cellular uplink communication may be protected. However due to the non-viability of the solution vis-a-vis practical implementation, a heuristic algorithm is proposed that builds upon the power allocation algorithm proposed for the overlay mode in order to find a feasible solution

In [58] a distributed resource and power allocation algorithm is proposed where the resource allocation problem is formulated as a mixed strategy non-cooperative game in which every user equipment in a cell is a player. Then a distributed spectral and energy efficient channel and power allocation algorithm is designed by minimizing power consumption and interference levels. The proposed method provides QoS guarantees for both D2D and cellular users.

**Table 6: Metric evaluation (Game theory-based resource allocation algorithms for D2D communication: Multiple D2D pair per channel)**

| References | Inband/ Outband | Underlay/ overlay | Uplink/ Downlink | D2D pairs per channel | Inter channel interference | Inter cellular interference | Channel sharing mode | Operational Design | Unicast/ Multicast | QoS Guarantees |
|---|---|---|---|---|---|---|---|---|---|---|
| [98] | Inband | Underlay | Downlink | Multiple | No | No | Many to one | Distributed | Unicast | CU |
| [99] | Inband | Underlay | Uplink | Multiple | No | No | Many to one | Distributed | Unicast | D2D |
| [100] | Inband | Underlay | Uplink | Multiple | No | No | Many to Many | Distributed Channel allocation, Centralized Power allocation | Multicast | CU and D2D |
| [101] | Inband | Underlay | Downlink | Multiple | No | Yes | Many to many | Distributed | Unicast | Both CU and D2D |
| [102] | Inband, outband | Underlay (for inband) | Uplink | Multiple | No | No | Many to one | Distributed | Unicast | No-Guarantees |
| [103] | Inband | Underlay | Downlink Uplink | Multiple | No | No | Many to one | Distributed | Unicast | CU |
| [104] | Inband | Underlay | Uplink | multiple | No | No | Many to one | Distributed | Unicast | CU |
| [105] | Inband | Overlay, underlay | Uplink | Multiple | No | Yes | -- | Hybrid | Unicast | CU |
| [58] | Inband | Underlay | Uplink | Multiple | No | No | Many to Many | Distributed | Unicast | CU and D2D |

## D. MACHINE LEARINGING BASED ALGORITHMS

Since the formulation of resource sharing problem contains binary resource allocation parameters, it ends up as a Non-Convex Mixed Integer Non-Linear Programming Problem (MINLP). Therefore, the conventional algorithms cannot achieve a globally optimal solution in realistic time frame. To overcome the shortcomings of conventional resource allocation algorithms, machine learning based algorithms are developed to solve the resource allocation problem. They employ an appropriate hierarchical structure of artificial neural networks that adopts a nonlinear method in order to effectively solve the problem of resource allocation in D2D communication networks. Fig. 4 presents a general frame work of machine learning based resource allocation for D2D communication.

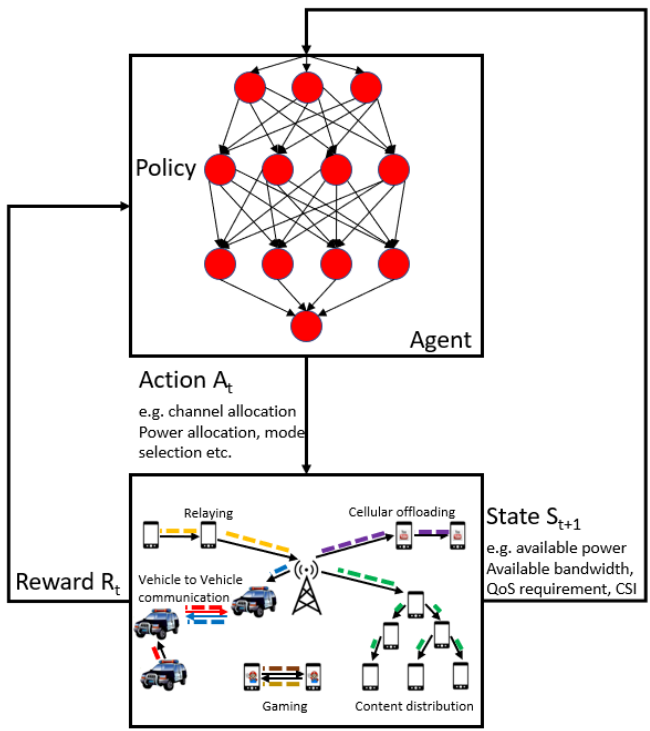

**Fig 4. General representation of machine learning based resource allocation for D2D communication.**

### 1) SINGLE D2D PAIR/GROUP PER CHANNEL

A deep Q-network based joint resource allocation and user association algorithm for ultra-dense D2D enabled networks is proposed in [106]. The proposed algorithm employs centralized deep reinforcement learning for resource allocation and user association of D2D pairs located in the overlapping regions of multiple small base stations. In this regard the central controller is considered as agent which interacts with the environment in order to carve out an optimal resource allocation and user association strategy for maximizing the system sum rate. The stated objective is achieved by first defining state, action and reward functions. Then a deep Q-network based design is proposed to achieve near optimal results. By employing deep reinforcement learning the proposed algorithm successfully avoids the constraint of availability of perfect CSI which is hard to satisfy in practical wireless communication scenarios. However, the proposed algorithm does not provide QoS guarantees for D2D and cellular users.

In [107], aiming at maximizing energy efficiency, the authors propose a hybrid multicast D2D clustering algorithm and a Q-Learning based joint subcarrier assignment and power control algorithm for multicast D2D communication underlaying cellular networks. First, keeping in view user preferences and communication reliability, an unsupervised learning-based clustering algorithm is proposed to form D2D clusters. Then, the joint subcarrier and power allocation problem is transformed into a mixed integer programming problem and a Q-learning and Lagrange dual decomposition-based solution is proposed for joint subcarrier assignment and power allocation to D2D users.

The authors in [108] propose a double deep Q-network based spectrum access algorithm that enables D2D users to learn optimal channel access strategies autonomously and maximize the sum rate without any prior information. Moreover, in order to deal with the unfair and stochastic resource allocation between D2D pairs the objective function is reformulated for ensuring both sum rate maximization and fairness among D2D users. Thus, the proposed method not only achieves optimal performance in terms of sum rate but also offers improved fairness in the resource allocation process.

The algorithm proposed in [109] combines the heuristic equally reduced power (ERP) method and deep neural network based method in order to improve the probability of achieving QoS guarantees in the form of minimum data rate for D2D users and maximum interference caused by D2D transmitters to the BS. The proposed method achieves near optimal sum rate while achieving low computation time. Moreover, minimum data rate is guaranteed for D2D communication. However, no such guarantees i.e., in terms of minimum achievable data rate, are provided for cellular users

Aimed at maximizing system level energy efficiency, the authors in [110] propose a resource allocation algorithm for D2D underlaid energy harvesting Internet of Things (IoT) networks. The formulated non-convex MINLP problem is solved by decomposition into three subproblems. Firstly, joint channel assignment and power splitting subproblems are addressed through a reinforcement learning based Q-learning algorithm. Then the power control subproblem is solved though a conventional convex optimization algorithm based on Dinkelbach and majorization-minimization methods.

**Table 7: Metric evaluation (Machine learning based resource allocation algorithms for D2D communication: One D2D pair per channel)**

| Reference | Inband/ Outband | Underlay/ overlay | Uplink/ Downlink | D2D pairs per channel | Inter channel interference | Inter cellular interference | Channel sharing mode | Operational Design | Unicast/ Multicast | QoS Guarantees |
|---|---|---|---|---|---|---|---|---|---|---|
| [106] | Inband | Underlay | downlink | One | No | Yes | One to One | Centralized | Unicast | None |
| [107] | Inband | Underlay | Uplink | One | No | No | One to one | Centralized | Multicast | CU |
| [108] | Inband | Underlay | Uplink | One | No | No | One to One | Centralized | Unicast | CU |
| [109] | Inband | Underlay | Uplink | One | No | No | One to one | Centralized | Unicast | D2D |
| [110] | Inband | Underlay (D2D underlaying IoT networks) | downlink | One | No | No | One to many | Centralized | Unicast | Both D2D and IoT users |

## 2) MULTIPLE D2D PAIR/GROUP PER CHANNEL

In [111], the Base Station (BS) acts as an agent that interacts with the environment at discrete times t. At each time t it observes state S(t) from the state space and implements an action from the action state in order to select a set of transmission powers based on a decision policy. At any time instant t, the input of the agent (i.e., BS) in the deep neural network is the state S(t) which is a D × D matrix whose elements are the normalized channel gains between a given pair of nodes. Whereas D is the number of transmitter-receiver pairs in a subchannel. The output is the action space that consists of N actions one of which is selected by the BS according to the decision policy while it is in state S(t). The proposed algorithm avoids the complexity of transmission power optimization in underlay D2D networks which is a mostly formulated as NP-hard combinatorial optimization problem.

Aimed at energy efficiency maximization in the long term, a Markov Decision Process (MDP) based problem is formulated in [112] in which users can dynamically switch between D2D mode and the conventional cellular mode. Then a deep reinforcement learning based algorithm referred to as deep deterministic policy gradient is employed to solve the formulated MDP problem. The setup of proposed algorithm comprises an actor network and a critic network, where the former generates deterministic actions using deterministic policy gradient method and the later assesses the performance of former using function-based Q networks. The proposed method is shown to achieve improved convergence and energy efficiency in a D2D based heterogeneous networks.

In [113], the resource allocation problem is modeled as a random non-cooperative game having multiple players in the form of D2D pairs. Each player acts as a learning agent who learns its best strategy using local information. To address the formulated problem, a double deep Q-Network based solution is proposed in which D2D users are not required to coordinate message exchanges with others users in the multi-user environment, thus enabling the system to reach its optimal communication state. The proposed algorithm reduces communication overhead by avoiding redundant message exchanges. On the downside, the QoS guarantees are provided only for cellular users.

Aimed at carrying out resource allocation with speedy convergence, the authors in [114] proposed a resource allocation approach based on game theory and deep reinforcement learning in a multiagent scenario. In particular, a critical Stackelberg Q-value is defined which is used to control the learning direction. The Q-value is determined based on the equilibrium attained in the considered Stackelberg game. With the guidance of the Stackelberg equilibrium, the proposed approach achieves faster convergence as compared to the general multi agent reinforcement methods and thus exhibits superior performance in in the management of network dynamics.

Aimed at satisfying the latency constraint for Vehicle to Vehicle(V2V) links, the author in [115], propose a multi-agent deep reinforcement learning based algorithm where each V2V link operates as an agent that interacts with the environment in order to make optimal decisions on transmission power and resource allocation. The decisions are made based on the observed environment and a reward function which depends on the latency of V2V links and the capacities of V2V and Vehicle to Infrastructure(V2I) links. In general, the proposed method achieves a balance between the latency requirements of V2V links and the interference from V2V transmitter to V2I receiver. In [115] only unicast scenario is considered. The authors further improve their work in [116] by incorporating broadcast scenarios where each receiver rebroadcasts the received message to improve reliability. However, rebroadcasting in a dense environment causes broadcast storms. This problem is addressed by allowing vehicles to rebroadcast a subset of the received messages such that more vehicle can receive the message within the latency constraint.

In [22], the authors proposed an energy-efficient autonomous channel allocation algorithm for D2D communication. The proposed algorithm employs a multiagent deep reinforcement learning based strategy where each D2D pair learns to autonomously select a channel based on the Q-values exchanged among neighboring D2D agents. The proposed method incurs higher communication overhead when compared with independent learning. However, on the upside, it achieves higher system throughput and speedy convergence.

## III. LIMITATIONS OF THE BOTTOM-LINE TECHNIQUES

In this section we identify the limitations of the bottom-line techniques adopted by the resource allocation algorithms for D2D communication (see Fig.5).

### A. LIMITATIONS OF THE CONVENTIONAL RESOURCE ALLOCATION ALGORITHMS

Following are some of the limitations of the conventional resource allocation algorithms

- Most of the conventional approaches necessitate perfect information about the wireless environment, such as instantaneous CSI and accurate channel models. However, due to the exploding growth in the number of connected devices, massive heterogeneity and high network density, acquiring such precise information will be very hard in next generation networks if not impossible.
- Conventional resource allocation algorithms are highly dependent on the underlaying system. Rapid variations in the system or wireless environment results in their inaccurate response. For proper functioning, they require reconfiguration in order to reflect the updated system settings. Nevertheless, contemporary wireless systems such as vehicular networks, drone swarms, railway networks etc., require support for highly dynamic systems characterized by rapidly changing

**Table 8: Metric evaluation (Machine learning based resource allocation algorithms for D2D communication: Multiple D2D pair per channel)**

| Reference | Inband/ Outband | Underlay/ overlay | Uplink/ Downlink | D2D pairs per channel | Inter channel interference | Inter cellular interference | Channel sharing mode | Operational Design | Unicast/ Multicast | QoS Guarantees |
|---|---|---|---|---|---|---|---|---|---|---|
| [111] | Inband | Underlay | Uplink | Multiple | No | No | Many to One | Centralized | Unicast | none |
| [112] | Inband | Overlay | Uplink | Multiple | No | Yes | Many to One | Centralized | Unicast | CU and D2D |
| [113] | Inband | Underlay | Uplink | Multiple | No | No | Many to One | Distributed | Unicast | CU |
| [114] | Inband | Underlay | Uplink | Multiple | No | No | Many to One | Distributed | Unicast | No guarantees in terms of SINR or data rate |
| [115] | Inband | Underlay | Uplink | Multiple | No | No | Many to One | Distributed | Can be unicast or multicast | Only V2V |
| [116] | Inband | Underlay | Uplink | Multiple | No | No | Many to One | Distributed | Unicast, broadcast | Only V2V |
| [22] | Inband | Underlay | Uplink | Multiple | No | No | Many to One | Distributed | Unicast | CU and D2D |

environment. This renders the conventional algorithms unworkable for such setups

- Most of the conventional resource allocation algorithms require intensive computation and sustain significant time delays. On one hand, longer time delays make them unfeasible for most of the evolving delay-sensitive applications e.g., autonomous vehicle communication and drone swarms etc., whereas on the other hand their lack of scalability in terms of computational complexity makes them unsuitable for large-scale wireless systems. Another major drawback of the conventional algorithms that stems out of their computational expensiveness, is that they can be implemented only in systems with high end computational capabilities. The miniature self-powered devices, which may form a considerable part of the future networks, might not be able to support such computationally intensive algorithms.
- Resource allocation optimization problems in wireless environments are usually complex and non-convex. Therefore, employing conventional optimization techniques to solve them will likely achieve local solutions instead of global optimal solutions.

### B. LIMITATIONS OF NOMA BASED RESOURCE ALLOCATION ALGORITHMS

Following are some of the limitations of the NOMA based resource allocation algorithms.

- Before decoding its own signal each user in a NOMA based setting is required to decode the signals from all other users in its group having worse channel gains. This leads to added complexity at the receiver and higher energy consumption as compared to orthogonal multiple access-based techniques.
- In the event of an error in successive interference cancellation at a user, the decoding of the signals from all the following users is likely to be erroneous. This implies keeping the number of users in each NOMA group reasonably small in order to diminish the effect of error propagation.
- In order to obtain the benefits of power-domain multiplexing, a significant difference in the channel gains of the weak and strong users is essential. This limits the number of user pairs in a NOMA group, which in turn decreases the sum rate gain of NOMA.
- NOMA introduces added communication overhead as every user is required to send its channel gain related statistics to the base station.
- NOMA is inherently sensitive to the uncertainty in the measurement of this gain.

### C. LIMITATIONS OF GAME THEORY BASED RESOURCE ALLOCATION ALGORITHMS

Following are some of the limitations of the game theory-based resource allocation algorithms.

- Game theory-based resource allocation is not suitable for networks with massively heterogeneous user devices and system architecture. In particular, for obtaining Nash Equilibrium it is assumed that all players are homogeneous with perfect network information and equal capabilities. Nonetheless, this is not true for modern wireless networks which are characterized by massively heterogeneous network entities with regards to computational, physical and communication capabilities. Moreover, obtaining perfect network knowledge is extremely exhaustive if not impossible.
- The computational complexity of game theory-based resource allocation methods and the volume of messages exchanged among contending players are directly proportional to the number of players. Thus, this inherent complexity coupled with the future expectancy of large-scale networks in terms of the number of user devices and network entities (e.g., Access Points, Base stations) makes the game theory-based algorithms prone to failure. In particular, exchange and update of the huge volumes of data and signaling among a huge number of players will generate unmanageable overhead besides drastic increase in computational complexity, delay, energy consumption and memory utilization of the players.

### D. LIMITATIONS OF MACHINE LEARNING BASED RESOURCE ALLOCATION ALGORITHMS

Following are some of the limitations of the machine learning based resource allocation algorithms.

- Supervised learning-based resource allocation requires collecting labelled data for training the deep neural network. Therefore, the optimal resource allocation policy must be found for a huge number of channel realizations which is an arduous task particularly for complex scenarios where large numbers of nodes are involved. Moreover, flexibility is another issue as entirely new data label must be generated in case of model changes such as in case of addition of a new constraint.
- Machine learning based algorithms need to be robust to the challenges presented by new D2D paradigms. For instance, mmWave has been identified as a potential area for D2D communication. However current machine learning algorithms do not consider the adversarial mmWave environment, such as path blocking and spatial transmissions coming from beam forming.
- Moreover, machine learning algorithms necessitate an incredible offline-learning stage which makes them unsuitable for future mmWave applications in 5G/6G.

P a g e 23 | 29

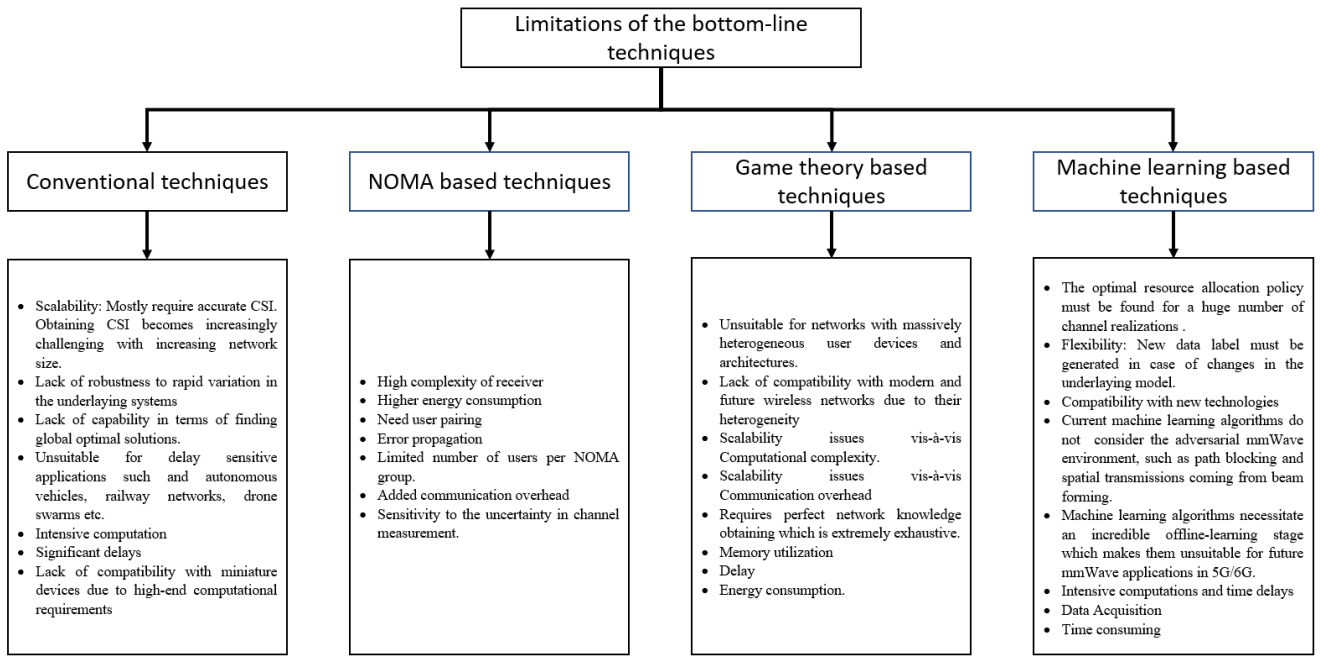

Fig 5. Limitations of underlaying resource allocation techniques

- Finally, reinforcement learning based methods typically function with settings that stay intact during learning. As a result, they may present inadequacies when dealing with dynamic scenarios where the state action spaces, state transition function and reward function change over time.

## IV. OPEN RESEARCH ISSUES

### A. CHANNEL STATE INFORMATION

Most of the existing literate assume that the perfect global or local channel state information is known apriori. Nonetheless estimation of CSI is not trivial as it requires frequent updates due to the dynamic nature of the wireless radio channel. Moreover, instantaneous CSI between all communicating D2D pairs (all transmitters and receivers in general) is required in order to make appropriate scheduling decisions. Therefore, as the number of the D2D users increases the communication overhead for CSI reporting increases too which may turn an otherwise feasible solution into a non-feasible solution due to intolerable overhead incurred for CSI reporting. Therefore, there is a need to study the scalability of the proposed resource allocation algorithms from the perspective of CSI.

### B. RELIABILITY AND LATENCY

Many applications of D2D communication have stringent reliability and latency requirements. Such applications include but are not limited to industrial automation and safety related services in V2V communication which are aimed at minimizing the probability of accidents. Nonetheless, the prevailing resource management algorithms for D2D communication seldom consider the latency requirement. Moreover, most of the prevalent resource management methods transform the reliability constraint directly into the target SINR ensuring that the achieved SINR is higher than the target SINR. However, the target SINR that guarantees reliability is different for modulation and coding schemes. Moreover, the existing algorithms consider only a snapshot of the system instead of addressing the dynamic behavior of the overall system as a stochastic process over time. Keeping in view the above-mentioned limitations of the existing work, there is a need for further research vis-à-vis latency and reliability. Methods that can be efficiently deployed in practical environments must be investigated.

### C. ACCESS RATE MAXIMIZATION

Most of the existing literature focus on sum rate maximization, outage probability minimization etc. for a limited number of D2D pairs/groups per channel. On the other hand, Access rate maximization is mostly ignored. Maximizing access rate per channel promises many benefits, such as higher overall sum rate due to admission of more D2D users, simultaneous serviceability for higher number of nodes and less power consumption per node. Therefore, there is a need to study resource allocation from the perspective of access rate maximization.

### D. INTERCELLULAR INTERFERENCE

Most of the existing literature assume single cell environment and therefore consider only intra cell interference. However, in practical scenarios, cells overlap and therefore cause intercellular interference. Intercellular interference degrades the performance of the edge users. Therefore, it is important to further explore the effect of inter cellular interference on resource allocation in multi cellular scenarios.

### E. ARCHITECTURE

Very few studies have explained the required architecture for supporting D2D communications underlaying cellular networks[117, 118]. Further research is required to study the ability of the present centralized cellular architecture in dealing with D2D procedures such as resource allocation, connection setup, interference control, security etc. Moreover, D2D communication should be studied in the more complex scenarios such as HetNets.

### F. FAIRNESS

Fairness refers to allocation of fair share of the available resources to the users. It is significant issue in D2D communication paradigm. However, the existing research on resource allocation for D2D communication focuses mainly on optimizing the overall throughput whereas the fairness aspect is mostly ignored. In principle, fairness and maximum throughput are inversely proportional. Optimizing the system throughput leads to reduction in the system fairness as demonstrated in [54, 117, 119-122]. On the contrary the system throughput will reduce if we try to maximize the system fairness. Keeping in view the above discussion, there is room for further research to address fairness problem while ensuring acceptable throughput levels

### G. COOPERATIVE COMMUNICATION

Cooperative communication is an efficient communication paradigm that can be employed for improving the performance of D2D underlaid cellular networks in terms of throughput and coverage maximization and interference minimization. In this regard, further research is required to enable the D2D users communicate in cooperative manner and optimize the available resources.

### H. MODELING OPTIMIZED PROBLEM FORMULATION

As the communication channel is dynamic with multiple constraints, the problem formulation for resource allocation is mostly concave and non-linear which makes it intractable and extremely complex to solve. The conventional solutions do not produce globally optimal results. Therefore, there is a need further explore areas such as artificial intelligence for improved resource allocation with limited complexity.

## V. CONCLUSION

In this paper we presented a comprehensive survey of state-of-the-art D2D communication algorithms, particularly from resource allocation perspective. The surveyed algorithms have been evaluated based on a set of metrics which constitute the elementary feature of a resource allocation algorithm for D2D communication. Additionally, the methodology of each algorithm is explained briefly for better understanding of the readers. The surveyed algorithms are categorized into four classes based on their bottom-line technology i.e., conventional optimization based, Non-Orthogonal-Multiple-Access (NOMA) based, game theory based and machine learning based techniques. Finally, we presented our view on the open challenges in the field of resource allocation for D2D communication. In the future, we plan to utilize the knowledge founded upon this survey to pursue novel solutions for those open challenges indicated in this survey.